%\magnification=\magstep1
%\hsize=140mm
%\vsize=180mm
%\hoffset=00mm
%\voffset=00mm
%\parskip=10pt
%\parindent=15pt
%\pageno=1
%\headline={\ifnum\pageno>1 \hss \number\pageno\  \hss \else\hfill \fi}
%\nopagenumbers
\vglue 12mm

\vskip 100pt
%----------------------------------------------------------------------------
\centerline{\bf \
{MIURA-LIKE FREE FIELD REALIZATION OF FERMIONIC CASIMIR ${\cal WB}_{_3}$
                                   ALGEBRAS\footnote{$^{^{_{\dagger}}}$}
{\it{This article is dedicated to the $225^{th}$ anniversary of the
Istanbul Technical University.}}
 }}
\vskip 5mm
\centerline{H.\ T.\ \"OZER \footnote{$^{^{_{\dagger \dagger}}}$}
{e-mail\ :\ ozert @ itu.edu.tr }}

\vskip 5mm
\centerline{\it {Physics Department,\ Faculty of Science and Letters,
Istanbul Technical University
,}}
\centerline{\it { 80626,\ Maslak,\ Istanbul,\ Turkey }}
\vskip 15mm
\noindent
Starting from the well-known quantum Miura-like transformation for the non
simply-laced Lie algebras ${\cal {B}}$${_{N}}$,\ we give an explicit
construction of the Casimir ${WB}_{_{3}}$ algebras .\ We reserve
the notation ${\cal{WB}}$$_{_N} $  for the Casimir ${\cal {W}}$ algebras
of type ${\cal {W}}$${{(2,4,6,\cdots,2 N,N+{{1}\over{2}})}}$\ which
contains one fermionic field. \ It is seen that ${WB}_{_{3}}$ algebra 
is closed and associative for all values of the central charge c.

\vskip 30mm

\par\vfill\eject

%----------------------------------------------------------------------------

\noindent{\bf{1.\ Introduction}}
\vskip5mm
\noindent  Extended (super) conformal algebras which consist of the
Virasoro algebra and with additional higher spin fields play an important
role  in our understanding the structure of two-dimensional (super)
conformal field theories\ (see e.g. Refs.\ 1-4).\ The idea to extend the
Virosoro algebra with the inroduction of higher integer and a half integer
conformal spin generators is also seem to be relevant in two-dimensional
conformal field theories (see Refs.1,5 and referances).
\ It is known that a Casimir algebra $^{5}$ corresponding
to a simple Lie algebra $\cal{L_{_N}}$ can be denoted by ${\cal{WL_{_N}}}^{4}$.\
Therefore Casimir ${\cal {WB}}$$_{_N}$ algebras based on non-simply laced
Lie algebra ${\cal {B}}$$_{_N}$.\ With the above identification,\
one says that  ${\cal {WB}}$$_{_1}$ algebra is the simplest
superconformal algebra.\ A construction of this type algebras was presented
first by Fateev and Luk'yanov $^{2}$.\ This algebra coincides with Neveu-
Schwarz algebra ($\cal{N}$=1 super Virasoro algebra).\ Then Figueroa-O'Farrill
at al.  have proven that ${\cal {WB}}$$_{_2}$ algebra are associative for
all values of central charge c $^{6}$.\ And also Ahn shown an explicit
construction of the Casimir ${\cal {WB}}$$_{_2}$ algebra $^{7}$ from a
fermion proposed by Watts in Ref.8.

In this paper we examine a Casimir algebra for a non simply-laced Lie algebras
$\cal{L_{_N}}=\cal{B_{_N}}$ ,\ which are called Casimir ${\cal {WB}}$$_{_N}$
algebras,\ for $N=3$,\ and also show that ${\cal {WB}}$$_{_3}$ algebra is closed
and associative for all values of c.\ We must emphasize here that we reserve
the notation  ${\cal {WB}}_{_N} $ for Casimir ${\cal {W}}$ algebras  of
type ${\cal {W}}$${{(2,4,6,\cdots,2 N+{{1}\over {2}})}}$ which contains
one fermionic field.\ Since there were three different $\cal{W}$ algebras
apart from our notation ${\cal {WB}}_{_N} $ algebra type
consistent for generic central charge$^{9,10}$i.e:

$\bullet$  The bosonic projection of the super Virasoro algebra $^{11}$.

$\bullet$  The Casimir algebra of ${\cal {B}}$ $_{3} $  ,obtained i.e.
by Hamiltonian reduction  for the Lie algebras ${\cal {B}}$ $_{_3}$ $^{11}$.

$\bullet$  The algebra described in Ref.12 .\ This algebra can be obtained
from a coset construction $^{4}$ or as a "unifying algebra"$^{13}$.

This paper is organized as follows.\ In section 2,\ we give a basis for
Casimir ${\cal {WB}}$$_{_N}$ algebra by using the well-known Miura-like
transformation with free massless bosonic fields and a free fermion field
$^{1,2}$.\
In section 3,\ we constructed a free field realization of the fermionic
Casimir ${\cal {WB}}$$_{_3}$ algebra by calculating explicitly all
nontrivial OPEs among primary fields.\ All results were given
by the shorthand notation.\ In section 4,\ we constructed a vertex
operator extension of the Casimir ${\cal {WB}}$$_{_N}$ algebras by
calculating explicitly nontrivial OPEs only between fermionic primary
field and vertex operators which are of two forms.\ The calculations of
OPEs have been done with the help of Mathematica Package OPEDefs.m of
Thielemans $^{14}$ under the {\bf{Mathematica$^{TM}$}} in Ref.15.

%\end
\vskip 10mm
%----------------------------------------------------------------------------
\noindent{\bf{ 2. \ The Casimir ${\cal {WB}}$$_{_N} $ Algebra Basis}}
\vskip 5mm

\noindent In this section we give a primary basis for the Casimir
${\cal {WB}}$$_{_N} $ algebras from  the free massless bosonic fields
and a free fermion field realization point of view $^{1,2,8}$.

\par The  Casimir ${\cal {WB}}$$_{_N}\  $  algebra generated by a set of
chiral currents $\ {U_{_{2 k}} (z)}\ $,\ of conformal dimension ${2 k}$~$~
(k=1 , \cdots , N) $,together with a fermionic field
$U_{_{N+{{1}\over{2}}}}(z)\ $, \ of conformal dimension $_N+{{1}\over{2}}$.\
The folloving Miura-like transformation
$$
R_{_N} (z)=
: \prod_{j=1}^N (\alpha_{_0} \partial_{_z}- h_{_j}(z) ):,
\eqno(2.1)
$$
and a free fermion field $b(z)$,\ of conformal dimension ${{1}\over{2}}$,\
give the fermionic field
$$
U_{_{N+{{1}\over{2}}}}(z)\ = \ R_{_N} (z)\ . \ b(z)
\eqno(2.2)
$$
of the Casimir ${\cal {WB}}$$_{_N} $ Algebra .
By taking OPE of
$d_{_{N+{{1}\over{2}}}}(z)$ with itself
one generates a set of
fields $\ {U_{_{2 k}} (z)}\ (k=1 , \cdots , N) $ above mentined.
In the Miura-like transformation,\ the symbol $ : ~* ~:$  shows
the normal ordering,\ and $\alpha_{_0}$ is a free parameter.
\ $h_{_j}(z)=i \mu_{_j}\partial \varphi(z) $ has $ N $ component
which are Feigin Fuchs-type of free massless scalar fields,\
here,\ ${\mu}_i $'s, $(i=1 , \cdots , N) $ are the weights
of the fundamental (vector) representation of  $B_{_{_N}} $,~satisfying
~${\mu}_i .{\mu}_j=\delta_{_{ij}} $.  \ The simple roots of  $B_{_{_N}} $
are given by $ {\alpha}_i={\mu}_i-{\mu}_{i+1} $, $(i=1 , \cdots , N-1) $
and $ {\alpha}_{_N}={\mu}_{_N} $ .\ A free scalar bosonic field
$ \varphi (z) $ and  a free fermionic field  $ b(z) $
are  single-valued function on the complex plane and its mode expansion
are given by
$$
i\,\partial{\varphi}(z)=\sum_{n \in Z} a_n z^{-n-1},\ \ \ \ \ \
b(z)=\sum_{n \in Z} b_n z^{-n-{{1}\over{2}}}.
\eqno(2.3)
$$
respectively.\ Canonical quantization give the commutator and
anti-commutator relations
$$
[a_m,a_n]=m \delta_{_{m+n,0}}\ , \ \ \ \ \ \ \
\{b_m,b_n\}=\delta_{_{mn}}\ ,
\eqno(2.4)
$$
and these  relations are equivalent to the contractions,\ as
$z_{12}=z_{1}-z_{2}$
$$
h_{_i}(\underline{z_1)\,h_{_j}(}z_2)={{\delta_{_{ij}}}\over {z_{12}^2}}
,\ \ \ \ \ \ \
b(\underline{z_1)\,b(}z_2)={{1}\over {z_{12}}}
\eqno(2.5)
$$
respectively.

\noindent First of all,\ in the example of ${\cal {WB}}$$_{_3} $ ,\
fermionic field
$U_{_{{{7}\over{2}}}}(z)\ $ can be obtained by expanding
$U_{_{N+{{1}\over{2}}}}(z)$ for $N=3$,\ which is following

$$
U_{_{{{7}\over{2}}}}(z)\ =(h_{_1} h_{_2}b)(z)
-a_{_0} (h_{_1}b')(z)
-a_{_0} (h_{_2}b')(z)
-a_{_0} (h'_{_2}b)(z)
+a^2_{_0} h''_{_2}(z)
\eqno(2.6)
$$

\noindent Then,\ by taking OPE of $U_{_{{{7}\over{2}}}}(z)$ with itself,\
under the following normalization,

$$
Q(z)=\sqrt{{3675}\over{2\ (49+c)\ (21+4 c)}}\ U_{_{{{7}\over{2}}}}(z)
\eqno(2.7)
$$

\noindent gives

$$
Q(z_{_1}) Q(z_{_2}) =\
{{2c/7}\over{{z^7_{_{12}}}}} +
{{2 W_{2}}\over{{z^5_{_{12}}}}} +
{{W_{2}'}\over{{z^4_{_{12}}}}} +
{{S_{4}}\over{{z^3_{_{12}}}}} +
{{S_{4}'}\over{{z^2_{_{12}}}}} +
{{S_{6}}\over{{z_{_{12}}}}}
\eqno(2.8)
$$

\noindent where

$$
W_{_2}(z)=\ {{1}\over{2}}
\sum_{i=1}^N\,\Big{(}(h_{_i}\,h_{_i})(z)+
\alpha_{_0}\,\big{(} i-N-{{1}\over{2}}\big{)} {h'_{_i}}(z) \Big{)}+
(b'b)(z)
\eqno(2.9)
$$

\noindent We observe that $ W_{_2} (z) \equiv U_{_2}(z)\equiv T(z) $ has spin-2,\ which
is called the stress-energy tensor.\ The standard OPE of $ T(z) $ with
itself is

$$ T(z_{1}) T(z_{_2})={{c/2} \over z^4_{_{12}}} +
{{2\  T} \over z^2_{_{12}}}+{{\partial{T}} \over z_{_{12}}}
\eqno(2.10)
$$
where the central charge, \ for  $B_{_{N}} $,\ is given by
$$
c=(N+{{1}\over{2}})\ (1-2 N(2 N-1) {\alpha_0}^2).
\eqno(2.11)
$$

\noindent We can also write  one of the two  statements in (8), i.e.
$S_{_4}(z)$ ,since  $S_{_6}(z)$ is too large,\ which is given by

$$
S_{_4}(z)=\,(1-2 \alpha^2_{_0})\,\Big{(}
\sum_{i<j}^3\,(h_{_i}\,h_{_i}\,h_{_j}\,h_{_j})(z)
+\sum_{i=1}^3\,(h_{_i}\,h_{_i} b'\,b)(z)
$$
$$
+\alpha_{_0}\,\sum_{i<j}^3\,(2\,j-7)\,(h_{_i}\,h_{_i}\,h'_{_j})(z)
-2\,\alpha_{_0}\,\sum_{i<j}^3\,(h_{_i}\,h'_{_j}\,h_{_j})(z)
-\alpha_{_0}\,\sum_{i=1}^3\,(h_{_i}\,b''\,b)(z)+
$$
$$
+\alpha^2_{_0}\,\sum_{i<j}^3\,(7-2\,j)\,(h_{_i}\,h''_{_j})(z)
+\alpha_{_0}\,\sum_{i<j}^3\,(2\,i-5)\,(h'_{_i}\,h_{_j}\,h_{_j})(z)
+\alpha_{_0}\,\sum_{i<j}^3\,(2\,i-5)\,(h'_{_i}\,b'\,b)(z)
\eqno(2.12)
$$
$$
+\alpha^2_{_0}\,\sum_{i<j}^3\,(2\,i-5)\,(2\,j-7)\,(h'_{_i}\,h'_{_j})(z)
+\alpha^2_{_0}\,\sum_{i=1}^3\,(i-1)\,(i-4)\,(h'_{_i}\,h'_{_i})(z)
$$
$$
+{{1}\over{2}}\,\sum_{i<j}^3\,(1-2\,(i^2-6\,i+11)\,\alpha^2_{_0})\,(h''_{_i}\,h_{_i})(z)
-{{1}\over{6}}\,\sum_{i<j}^3\,(7-2\,i)\,(1-\,(i^2-7\,i+18)\,\alpha^2_{_0})\,\alpha_{_0}\,h'''_{_i}(z)
$$
$$
+{{1}\over{6}}\,(1-6\,\alpha^2_{_0})(b'''b)(z)\Big{)}
$$
We can decompose $S_{_4}(z)$ and $S_{_6}(z)$ to the fields
$W_{_{2k}}(z)$'s ,\ as follows

$$
S_{_4}(z)=
\alpha_{_1} W_{_4}(z) +
\alpha_{_2} (W_{_2}\,W_{_2})(z)+
\alpha_{_3} W''_{_2}(z)
\eqno(2.13)
$$
\noindent and
$$
S_{_6}(z)=
\alpha_{_5} W''_{_4}(z)+
\alpha_{_6} (W''_{_2}W_{_2})(z)+
\alpha_{_7} (W'_{_2}W'_{_2})(z)+
\alpha_{_8} W''''_{_2}(z)+
\alpha_{_9} (W_{_2}\,W_{_2}\,W_{_2})(z)
$$
$$
+\alpha_{_{10}} W_{_6}(z) +
\alpha_{_{11}} (W_{_2}\,W_{_4})(z)
\eqno(2.14)
$$

\noindent Notice that $W_{_4}(z)$ and $W_{_6}(z)$ are not a primary field
under the stress-energy tensor  $T(z)$ .\ Therefore we can give the
primary fields $U_{_{2 k}}(z)$ in the following normalizations

$$
U_{_4}(z)=\sqrt{{\left( 21 + 4\,c \right) \,\left( 22 + 5\,c \right) }\over
    {2\,\left( 19 + 6\,c \right) \,\left( 161 + 8\,c \right) }}\ W_{4}(z)
\equiv U(z)
\eqno(2.15)
$$
$$
U_{_6}(z) =\sqrt{{3\,\left( 24 + c \right) \,\left( -1 + 2\,c \right) \,
      \left( 21 + 4\,c \right) \,\left( 68 + 7\,c \right) }\over
    {5\,\left( -1 + c \right) \,\left( 19 + 6\,c \right) \,
      \left( 161 + 8\,c \right) \,\left( 403 + 22\,c \right) }}\ W_{6}(z)
\equiv R(z)
\eqno(2.16)
$$
\noindent respectively.\ Thus,\ above normalized fields leads to new
decompositions for  $S_{_4}(z)$ and $S_{_6}(z)$ .\ One obtains

$$
S_{_4}(z)=
a_{_1} U(z) +
a_{_2} (T\,T)(z)+
a_{_3} T''(z)
\eqno(2.17)
$$
\noindent and
$$
S_{_6}(z)=
a_{_5} U''(z)+
a_{_6} (T''\,T)(z)+
a_{_7} (T'\,T')(z)+
a_{_8} T''''(z)+
a_{_9} (T\,T\,T)(z)
$$
$$
+a_{_{10}} R(z) +
a_{_{11}} (T\,U)(z)
\eqno(2.18)
$$
\noindent where all $a_{_i}$'s are given by

$$\eqalign{
&a_{_1}^2={{2\,\,{{\cal {E}}_{_{1}}}\,{{\cal {E}}_{_{2}}}}\over{{{\cal {E}}_{_{3}}}\,{{\cal {E}}_{_{4}}}}}
\,,\,
a_{_2}={{37}\over{{{\cal {E}}_{_{4}}}}}
\,,\,
a_{_3}={3\over2}\,{{{{\cal {E}}_{_{5}}}}\over{{{\cal {E}}_{_{4}}}}}
\,,\,
a_{_4}={{{{\cal {E}}_{_{6}}}}\over{12\,{{\cal {E}}_{_{4}}}}}
\,,\,
a_{_5}^2={{25}\over{648}}\,{{{{\cal {E}}_{_{1}}}\,{{\cal {E}}_{_{2}}}\,{{{{\cal {E}}_{_{7}}}}^2}}\over{{{\cal {E}}_{_{3}}}\,{{\cal {E}}_{_{4}}}\,{{{{\cal {E}}_{_{8}}}}^2}}}
\,,\,
a_{_6}={5\over2}\,{{{{\cal {E}}_{_{9}}}}\over{{{\cal {E}}_{_{10}}}\,{{\cal {E}}_{_{11}}}\,{{\cal {E}}_{_{4}}}}}
\,,\,
a_{_7}={5\over2}\,{{{{\cal {E}}_{_{12}}}}\over{{{\cal {E}}_{_{10}}}\,{{\cal {E}}_{_{11}}}\,{{\cal {E}}_{_{4}}}}}
\cr
&a_{_8}={5\over{12}}\,{{{{\cal {E}}_{_{13}}}}\over{{{\cal {E}}_{_{10}}}\,{{\cal {E}}_{_{11}}}\,{{\cal {E}}_{_{4}}}}}
\,,\,
a_{_9}={{5\,\,{{\cal {E}}_{_{14}}}}\over{{{\cal {E}}_{_{10}}}\,{{\cal {E}}_{_{11}}}\,{{\cal {E}}_{_{4}}}}}
\,,\,
a_{_{10}}^2=
{5 \over 3} \,{{{ {\cal {E}}_{_{1}}}\, \,{ {\cal {E}}_{_{2}}}\, \,{ {\cal {E}}_{_{15}}}\, \,{ {\cal {E}}_{_{16}}}}\over    { \,{ {\cal {E}}_{_{3}}}\, \,{ {\cal {E}}_{_{8}}}\, \,{ {\cal {E}}_{_{10}}}\, \,{ {\cal {E}}_{_{11}}}}}
\,,\,
a_{_{11}}^2={{1250}\over{9}}{{{{\cal {E}}_{_{1}}}\,{{\cal {E}}_{_{2}}}}\over{{{\cal {E}}_{_{3}}}\,{{\cal {E}}_{_{4}}}\,{{{{\cal {E}}_{_{8}}}}^2}}}
}
\eqno(2.19)
$$

\noindent where  we set

$$\eqalign{
&{\cal{E}}_{{1}}= 19\,+6\,c
\,,\,
{\cal{E}}_{{2}}=161\,+\,8\,c
\,,\,
{\cal{E}}_{{3}}= 21\,+\,4\,c
\,,\,
{\cal{E}}_{{4}}= 22\,+\,5\,c
\,,\,
{\cal{E}}_{{5}}= -3\,+\,c
\,,\,
{\cal{E}}_{{6}}=-49\,+\,4\,c
\,,\,
{\cal{E}}_{{7}}= 14\,+\,c
\cr
&{\cal{E}}_{{8}}= 24\,+\,c
\,,\,
{\cal{E}}_{{9}}= -793\,+\,103\,c\,+\,62\,c^2
\,,\,
{\cal{E}}_{{10}}=-1\,+\,2\,c
\,,\,
{\cal{E}}_{{11}}=68\,+\,7\,c
\,,\,
{\cal{E}}_{{12}}= 270\,+\,457\,c\,+\,52\,c^2
\cr
&{\cal{E}}_{{13}}=-27-246\,c\,+\,5\,c^2\,+\,2\,c^3
\,,\,
{\cal{E}}_{{14}}=41\,+\,270\,c
\,,\,
{\cal{E}}_{{15}}=-1\,+\,c
}
\eqno(2.20)
$$
\par\vfill\eject

%----------------------------------------------------------------------------
\vskip 5mm
\noindent{\bf{ 3. \  ${WB}_{_{3}}$ OPEs With Shorthand Notation}}
\vskip 5mm

In this section we present the results of our explicit calculations of
${\cal WB}_{_3}$ OPEs.\ In all these OPEs,\ the composite fields are
included only up to conformal spin 10 ,\ the higher order composite fields
neglected due to their formal complexity.\ In view of this shorthand
notation,\ we verified the OPEs of the complete Casimir ${\cal WB}_{_3}$
algebra

$$\eqalign{
& Q \, \star \, Q \,=
        \, {{2\,c }\over{7}}     \, I
\,\,+ \,\, {\cal{C}^{_U}_{_{QQ}}}\, U
\,\,+ \,\, {\cal{C}^{_R}_{_{QQ}}}\, R
\cr
& U \, \star \, Q \,=
        \, {\cal{C}^{_Q}_{_{UQ}}}\, Q
\cr
& U \, \star \, U \,=
        \, {{c}\over{4}}         \, I
\,\,+ \,\, {\cal{C}^{_U}_{_{UU}}}\, U
\,\,+ \,\, {\cal{C}^{_R}_{_{UU}}}\, R
\cr
& R \, \star \, Q \,=
        \, {\cal{C}^{_Q}_{_{RQ}}}\, Q
\,\,+ \,\,  \tilde {\cal{C}}^{_{(UQ)}}_{_{RQ}}\, (U\,Q)
\cr
& U \, \star \, R \,=
        \,  {\cal{C}^{_U}_{_{UR}}}\, U
\,\,+ \,\,  {\cal{C}^{_R}_{_{UR}}}\, R
\,\,+ \,\,  \tilde {\cal{C}}^{_{(UU)}}_{_{UR}}\, (U\,U)
\cr
& R \, \star \, R \,=
        \,  {{c}\over{6}}\, I
\,\,+ \,\,  {\cal{C}^{_U}_{_{RR}}}\, U
\,\,+ \,\,  {\cal{C}^{_R}_{_{RR}}}\, R
\,\,+ \,\,  {\tilde{\cal{C}}}^{_{(UU)}}_{_{RR}}\, (U\,U)
\,\,+ \,\,  {\tilde{\cal{C}}}^{_{(UR)}}_{_{RR}}\, (U\,R)
}
\eqno(3.1)
$$
\noindent Here we write down the expressions for the coefficients in the
above OPE's for ${\cal WB}_{_3}$ algebras.
$$\eqalign{
& \big{(}{\cal{C}^{_U}_{_{QQ}}}\big{)}^{_2}\,=\,a_{_1}^2
\,,\,\,\,
  \big{(}{\cal{C}^{_R}_{_{QQ}}}\big{)}^{_2}\,=\,a_{_{10}}^2
\,,\,\,\,
  \big{(}{\cal{C}^{_Q}_{_{UQ}}}\big{)}^{_2}\,=\,
{{49}\over{32}}{{ \,{ {\cal {E}}_{_{1}}}\, \,{ {\cal {E}}_{_{2}}}}\over { \,{ {\cal {E}}_{_{3}}}\, \,{ {\cal {E}}_{_{4}}}}}
\,,\,\,\,
  \big{(}{\cal{C}^{_U}_{_{UU}}}\big{)}^{_2}\,=\,
{{2\, \,{{{ {\cal {E}}_{_{17}}}}^2}}\over {{ {\cal {E}}_{_{1}}}\, \,{ {\cal {E}}_{_{2}}}\, \,{ {\cal {E}}_{_{3}}}\, \,{ {\cal {E}}_{_{4}}}}}
\cr
& \big{(}{\cal{C}^{_R}_{_{UU}}}\big{)}^{_2}\,=\,
{{80}\over{3}}{{ \,{{{ {\cal {E}}_{_{4}}}}^2}\, \,{{{ {\cal {E}}_{_{18}}}}^2}\, \,{ {\cal {E}}_{_{15}}}\, \,{ {\cal {E}}_{_{16}}}}\over    { \,{ {\cal {E}}_{_{1}}}\, \,{ {\cal {E}}_{_{2}}}\, \,{ {\cal {E}}_{_{3}}}\, \,{ {\cal {E}}_{_{8}}}\, \,{ {\cal {E}}_{_{10}}}\, \,{ {\cal {E}}_{_{11}}}}}
\,,\,\,\,\,\,
 \big{(}{\cal{C}^{_Q}_{_{RQ}}}\big{)}^{_2}\,=\,
{{245}\over{432}}{{ \,{ {\cal {E}}_{_{1}}}\, \,{ {\cal {E}}_{_{2}}}\, \,{ {\cal {E}}_{_{15}}}\, \,{ {\cal {E}}_{_{16}}}}\over    { \,{ {\cal {E}}_{_{3}}}\, \,{ {\cal {E}}_{_{8}}}\, \,{ {\cal {E}}_{_{10}}}\, \,{ {\cal {E}}_{_{11}}}}}
\,,\,\,\,\,\,
 \big{(}{{\tilde{\cal{C}}}^{_{(UQ)}}_{_{RQ}}}\big{)}^{_2}\,=\,
{{245}\over{6}}{{\,{ {\cal {E}}_{_{10}}}\, \,{ {\cal {E}}_{_{11}}}\, \,{ {\cal {E}}_{_{16}}}}\over{ \,{{{ {\cal {E}}_{_{18}}}}^2}\, \,{ {\cal {E}}_{_{4}}}\, \,{ {\cal {E}}_{_{8}}}\, \,{ {\cal {E}}_{_{15}}}}}
\cr
& \big{(}{\cal{C}^{_U}_{_{UR}}}\big{)}^{_2}\,=\,
{{320}\over{27}}{{\,{{{ {\cal {E}}_{_{4}}}}^2}\, \,{{{ {\cal {E}}_{_{18}}}}^2}\, \,{ {\cal {E}}_{_{15}}}\, \,{ {\cal {E}}_{_{16}}}}\over    { \,{ {\cal {E}}_{_{1}}}\, \,{ {\cal {E}}_{_{2}}}\, \,{ {\cal {E}}_{_{3}}}\, \,{ {\cal {E}}_{_{8}}}\, \,{ {\cal {E}}_{_{10}}}\, \,{ {\cal {E}}_{_{11}}}}}
\,,\,\,\,\,\,
 \big{(}{\cal{C}^{_R}_{_{UR}}}\big{)}^{_2}\,=\,
{{{{{ {\cal {E}}_{_{10}}}}^2}\, \,{{{ {\cal {E}}_{_{11}}}}^2}\, \,{{{ {\cal {E}}_{_{19}}}}^2}}\over    {18\, \,{ {\cal {E}}_{_{1}}}\, \,{ {\cal {E}}_{_{2}}}\, \,{ {\cal {E}}_{_{3}}}\, \,{ {\cal {E}}_{_{4}}}\, \,{{{ {\cal {E}}_{_{8}}}}^2}}}
\,,\,
 \big{(}{{\tilde{\cal{C}}}^{_{(UU)}}_{_{UR}}}\big{)}^{_2}\,=\,
{{4000}\over{3}}{ {{ {\cal {E}}_{_{10}}} \,{ {\cal {E}}_{_{11}}}}\over    {\,{ {\cal {E}}_{_{4}}}\, \,{ {\cal {E}}_{_{8}}}\, \,{ {\cal {E}}_{_{15}}}\, \,{ {\cal {E}}_{_{16}}}}}
\cr
& \big{(}{\cal{C}^{_U}_{_{RR}}}\big{)}^{_2}\,=\,
{2\over{81}}{{\,{{{ {\cal {E}}_{_{10}}}}^2}\, \,{{{ {\cal {E}}_{_{11}}}}^2}\, \,{{{ {\cal {E}}_{_{19}}}}^2}}\over    { \,{ {\cal {E}}_{_{1}}}\, \,{ {\cal {E}}_{_{2}}}\, \,{ {\cal {E}}_{_{3}}}\, \,{ {\cal {E}}_{_{4}}}\, \,{{{ {\cal {E}}_{_{8}}}}^2}}}
\,,\,\,\,
 \big{(}{\cal{C}^{_R}_{_{RR}}}\big{)}^{_2}\,=\,
{{80}\over{243}}{{ \,{{{ {\cal {E}}_{_{20}}}}^2}}\over    { \,{ {\cal {E}}_{_{1}}}\, \,{ {\cal {E}}_{_{2}}}\, \,{ {\cal {E}}_{_{3}}}\, \,{ {\cal {E}}_{_{10}}}\, \,{ {\cal {E}}_{_{11}}}\, \,{{{ {\cal {E}}_{_{8}}}}^3}\, \,      { {\cal {E}}_{_{15}}}\, \,{ {\cal {E}}_{_{16}}}}}
\cr
&  {\tilde{\cal{C}}}^{_{(UU)}}_{_{RR}}\,=\,
{{{ {\cal {E}}_{_{10}}}\, \,{ {\cal {E}}_{_{11}}}\, \,{ {\cal {E}}_{_{21}}}}\over    {3\, \,{ {\cal {E}}_{_{4}}}\, \,{ {\cal {E}}_{_{8}}}\, \,{ {\cal {E}}_{_{15}}}\, \,{ {\cal {E}}_{_{16}}}\, \,{ {\cal {E}}_{_{18}}}}}
\,,\,\,\,
 \big{(}{\tilde{\cal{C}}}^{_{(UR)}}_{_{RR}}\big{)}^{_2}\,=\,
{{120\, \,{{{ {\cal {E}}_{_{22}}}}^2}\, \,{ {\cal {E}}_{_{10}}}\, \,{ {\cal {E}}_{_{11}}}}\over    {{{{ {\cal {E}}_{_{18}}}}^2}\, \,{ {\cal {E}}_{_{4}}}\, \,{ {\cal {E}}_{_{8}}}\, \,{ {\cal {E}}_{_{15}}}\, \,{ {\cal {E}}_{_{16}}}}}
}
\eqno(3.2)
$$
\noindent where  we set
$$\eqalign{
&{\cal{E}}_{{16}}=403\,+\,22\,c
\,,\,
{\cal{E}}_{{17}}= 4214\,+\,627\,c\,+\,34\,c^2
\,,\,
{\cal{E}}_{{18}}=49\,+\,c
\,,\,
{\cal{E}}_{{19}}=812\,+\,13\,c
\,,\,
{\cal{E}}_{{20}}=-253941632
\cr
&-36692548\,c\,+\,37936207\,c^2\,+\,3693737\,c^3\,+\,97882\,c^4\,+\,104\,c^5
\,,\,
{\cal{E}}_{{21}}= 522592\,+\,25091\,c\,+\,442\,c^2
\cr
&{\cal{E}}_{{22}}=14\,+\,11\,c
}
\eqno(3.3)
$$

\vskip 5mm

\noindent{\bf{ 4.\ (OPEs) for Chiral Vertex Operators}}
\vskip 5mm
\noindent
We define two kinds of vertex operators which  corresponds to the long and
short roots of $B_{_N}$
$$
{\cal{V}}^l_{\beta}(z)=:\,e^{i\,\beta\,.\,\varphi(z)}\,:
\eqno(4.1)
$$
and
$$
{\cal{V}}^s_{\beta}(z)=:b(z)\,e^{i\,\beta\,.\,\varphi(z)}\,:
\eqno(4.2)
$$
Here a non-simple root $\beta$ including the long and short roots of
$B_{_{N}}$ ,\ is given by
$$
\beta=\sum_{i=1}^{_{N}} m_{_i} \alpha_{_i}
\eqno(4.3)
$$
and  The Fubini-Veneziano field $ \varphi(z)$,~which has conformal spin-0
$$
\varphi(z)=q-i p \ln z + i \sum_{_{n \ne 0}} {1 \over n} a_{_n} z^{-n}
\eqno(4.4)
$$
By using conformal spin-0  contraction $\varphi(\underline{z)\,\varphi(}w)=
-\ln \mid z-w \mid$,\ The standard OPEs are of two forms:
$$
{\cal{V}}^l_{_\beta}(z)\,{\cal{V}}^l_{_{\dot{\beta}}}(w)\,=( z-w)^{\beta\,\dot{\beta}}\,
:{\cal{V}}^l_{_\beta}(z)\,{\cal{V}}^l_{_{\dot{\beta}}}(w):
\eqno(4.5)
$$
and
$$
{\cal{V}}^s_{_\beta}(z)\,{\cal{V}}^s_{_{\dot{\beta}}}(w)\,={1\over{( z-w)}}
{\cal{V}}^l_{_\beta}(z)\,{\cal{V}}^l_{_{\dot{\beta}}}(w)
\eqno(4.6)
$$
We can tell that the operators ${\cal{V}}^l_{_\beta}(z)$ and  ${\cal{V}}^s_{_\beta}(z)$
carrie a root $\beta$ .\ From
$$
h_{_j}(z)\,{\cal{V}}^l_{\beta}(w)=
{{\theta_{_j}}\over {z-w}}  {\cal{V}}^l_{_{\beta}}(w)\,+\,\cdots
\eqno(4.5)
$$
and
$$
h_{_j}(z)\,{\cal{V}}^s_{\beta}(w)=
{{\theta_{_j}}\over {z-w}}  {\cal{V}}^s_{_{\beta}}(w)\,+\,\cdots
\eqno(4.6)
$$
where
$$
\theta_{_j}=
\theta_{_j}(\beta)
\equiv\,
(\beta,\mu_{_{j}})\,
\eqno(4.6)
$$
The OPEs with the stress-energy tensor $T(z)$ is
$$
T(z)\,{\cal{V}}^l_{\beta}(w)=
{h^l(\beta)\over {(z-w)^2}}  {\cal{V}}^l_{_{\beta}}(w) \,+\,
{({\eta_{_l}}^{\beta} {\cal{V}}^l_{_{\beta}})(w)\over {z-w}} \,+\,\cdots
\eqno(4.7)
$$
where~$({\eta_{_l}}^{\beta} {\cal{V}}^l_{_{\beta}})(z)$
\footnote{$^{\star}$}
{if we take $\beta=\,\alpha_{_i}$,~ a simple root,~ then ~
$({\eta_{_l}}^{\beta} {\cal{V}}^l_{_{\beta}})(z)
=\partial {\cal{V}}^l_{_{\beta}}(z)$
and
$({\eta_{_s}}^{\beta} {\cal{V}}^s_{_{\beta}})(z)
=\partial {\cal{V}}^s_{_{\beta}}(z)
.$}
,\ is given by
$$
({\eta_{_l}}^{\beta} {\cal{V}}^l_{_{\beta}})(z)=
\sum_{i}^N\,
\theta_{_i}\,
(h_{_i} {\cal{V}}^l_{_{\beta}}) (z)\,
\eqno(4.8)
$$
and similarly , for the short root
$$
T(z)\,{\cal{V}}^s_{\beta}(w)=
{h^s(\beta)\over {(z-w)^2}}  {\cal{V}}^s_{_{\beta}}(w) \,+\,
{({\eta_{_s}}^{\beta} {\cal{V}}^s_{_{\beta}})(w)\over {z-w}} \,+\,\cdots
\eqno(4.9)
$$
where~$({\eta_{_s}}^{\beta} {\cal{V}}^s_{_{\beta}})(z)^{~\star}$,\ is given by
$$
({\eta_{_s}}^{\beta} {\cal{V}}^s_{_{\beta}})(z)=
\sum_{i}^N\,
\theta_{_i}\,
(h_{_i} {\cal{V}}^s_{_{\beta}}) (z)+(\partial{b}\,{\cal{V}}^l_{_{\beta}})(z) \,
\eqno(4.10)
$$
Thus the vertex operators ${\cal{V}}^l_{\beta}(z)$ and ${\cal{V}}^s_{\beta}(z)$
are a conformal field of spin $h^l(\beta)$  and  $h^s(\beta)$
which are algebraic in $\beta$,\ are given by \ respectively
$$
h^l(\beta)= {{1}\over{2}}\,
\sum_{i}\big{\{}\,\theta^2_{_i}\,+\,
2\, \alpha_{_0}\,(N-i+{{1}\over{2}})\,\theta_{_i}\,\big{\}}
\eqno(4.11)
$$
and
$$
h^s(\beta)=h^l(\beta)\,+\,{{1}\over{2}}
\eqno(4.12)
$$
Another important OPEs  between the Fermionic field
$U_{_{N+{{1}\over{2}}}}(z)$ and the Vertex Operators have been calculated
explicitly ,\ the lower order singler terms are neglected due to their
formal complexity.\ We have
$$
U_{{N+{{1}\over{2}}}}(z)\,{{\cal{V}}^l_{\beta}}(w)=
{U^l_{{N+{{1}\over{2}}}}(\beta)\over {(z-w)^N}}  {\cal{V}}^s_{_{\beta}}(w) \,+\,\cdots
\eqno(4.13)
$$
and also
$$
U_{{N+{{1}\over{2}}}}(z)\,{{\cal{V}}^s_{\beta}}(w)=
{U^s_{{N+{{1}\over{2}}}}(\beta)\over {(z-w)^{N+1}}}  {\cal{V}}^l_{_{\beta}}(w) \,+\,\cdots
\eqno(4.14)
$$
where
$$
U^l_
{N+{{1}\over{2}}}(\beta)\,=\, (-1)^N\,\prod\limits_{i=1}^N
\Big(\theta_i\,+\,(N-i+1) \alpha_0\Big)
\eqno(4.15)
$$
and
$$
U^s_
{N+{{1}\over{2}}}(\beta)\,=\, (-1)^N\,\prod\limits_{i=1}^N
\Big(\theta_i\,+\,(N-i) \alpha_0\Big)
\eqno(4.16)
$$
%\par\vfill\eject
%----------------------------------------------------------------------------

\noindent{\bf Acknowledgments}
\vskip 5mm

\noindent The author would like to thank H.\ R.\ Karadayi for his valuable
discussions and excellent guidance throughout this research.
\vskip 2mm
\noindent It is a pleasure to thank R. Blumenhagen, H.G. Kausch, K.Thielemans,
A. Honecker, A. Sorin, G.M.T. Watts and  S. Belucci  for helpful
informations via e-mail communications.
\vskip 2mm
\noindent The calculation of OPE has been done with the help of Mathematica
Package OPEDefs.m $^{14}$.

\vskip 5mm

%----------------------------------------------------------------------------
\noindent{\bf References}

\vskip 5mm

\noindent 1. P.\ Bouwknegt and K.\ Schoutens, {\it{Phys. Rep}}. {\bf{223}} (1993) 183;

\noindent 2. V.A. Fateev and S.L. Luk'yanov, {\it{Sov. Sci. Rev. A Phys}}.15/2 (1990)

\noindent 3. A. B. Zamolodchikov, {\it{Theor. Math. Phys}}. {\bf{65}}, 1205 (1986).

\noindent 4. R.\ Blumenhagen,\ W.\ Eholzer, A.\ Honecker R.\ Hubel  and
K.\ Hornfeck,\ {\it{Int.\ J.\ Mod.\ Phys}}.{\bf{A10}}, 2367 (1995).

\noindent 5. F. Bais, P. Bouwknegt, M. Surridge and K. Schoutens,{\it{ Nucl. Phys}}.
{\bf{B304}} (1988) 348; 371.

\noindent 6. J.M. Figueroa-O'Farrill, S. Schrans and K. Thielemans, "On the
Casimir algebra of $B_2$", {\it{Phys. Lett}}. {\bf{263B}} (1991) 378.

\noindent 7. C.\ "Ahn c=5/2 Free Fermion Model of ${WB}_{_{2}}$ Algebra",
{\it{Int. J. Mod. Phys}}. {\bf{A 6}} (1991) 3467.

\noindent 8. G.M.T. Watts, "$\cal{WB}$ Algebra Representation Theory",
{\it{Nucl. Phys}}. {\bf{B339}} (1990) 177; "$\cal{W}$- Algebras and Coset Models",
{\it{Phys. Lett}}. {\bf{245B}} (1990)65, DAMTP-90-37.

\noindent 9. H.G. Kausch, private communication.

\noindent 10. R.\ Blumenhagen, private communication.

\noindent 11. H.G. Kausch and G.M.T. Watts, " A Study of $\cal{W}$ Algebras
Using Jacoby Identities", {\it{Nucl. Phys}}. {\bf{B354}},740 (1991).

\noindent 12. W.Eholzer, A.Honecker and R.Hubel, " How Complete is the
Classification of $\cal{W}$-Symmetries ?",\ hep-th/9302124 (a revised and
extended version of {\it{Phys. Lett}}. {\bf{B308}} (1993) p.42 )

\noindent 13. R.\ Blumenhagen,\ W.\ Eholzer, A.\ Honecker, K.\ Hornfeck
and R.\ Hubel, preprint BONN-TH-94-01,\ hep-th/9404113.

\noindent 14. K.\ Thielemans,"A  ${Mathematica^{TM}}$ package for computing
operator product expansions (OPEdefs 3.1)",\ Theoretical Phys.Group,\ Imperial
College,\ London(UK).

\noindent 15.\ S. Wolfram, ${Mathematica^{TM}}$, (Addison-Wesley,1990).

\end